\documentclass[useAMS,usenatbib]{mn2e}
\usepackage{subfigure}
\usepackage{xspace}
\usepackage{graphicx, amssymb}
\usepackage[fleqn]{amsmath}

\newcommand{\kelvin}{\,{\rm K}}
\newcommand{\s}{\,{\rm s}}

\newcommand{\cc}{\,{\rm cm}^{-3}}
\newcommand{\msun}{\,{\rm M}_{\odot}}

\newcommand{\kms}{\,\mathrm{km}\,\mathrm{s}^{-1}}
\newcommand{\au}{\,{\rm AU}}
\newcommand{\pc}{\,{\rm pc}}
\newcommand{\kpc}{\,{\rm kpc}}
\newcommand{\Mpc}{\,{\rm Mpc}}

\newcommand{\yr}{\,{\rm yr}}

\newcommand{\ev}{\,{\rm eV}}

\newcommand{\erg}{\,{\rm erg}}

\newcommand{\mh}{m_{\mathrm{H}}}
\newcommand{\htwo}{\mathrm{H}_2}
\newcommand{\hd}{\mathrm{HD}}
\newcommand{\deut}{\mathrm{D}}
\newcommand{\h}{\mathrm{H}}
\newcommand{\hplus}{\mathrm{H}^+}
\newcommand{\hminus}{\mathrm{H}^-}
\newcommand{\he}{\mathrm{He}}
\newcommand{\heplus}{\mathrm{He}^+}

\newcommand{\ucr}{{u}_{\rm \small CR}}
\newcommand{\ucrz}{{u}_{\rm \small CR}(z)}
\newcommand{\Dhubble}{D_{\mathrm{H}}}

\title[First stars under cosmic ray feedback]{The First Stars: formation under cosmic ray feedback}

\author[J.A.~Hummel et al.]{Jacob~A.~Hummel$^{1}$, Athena~Stacy$^{2}$ and Volker~Bromm$^{1}$\\
$^1$Department of Astronomy, The University of Texas at Austin, TX 78712, USA\\
$^2$Department of Astronomy, University of California, Berkeley, CA 94720, USA}

\begin{document}

%\date{Accepted 1988 December 15. Received 1988 December 14; in original form 1988 October 11}

%\pagerange{\pageref{firstpage}--\pageref{lastpage}} \pubyear{2002}

\maketitle

\begin{abstract}
We explore the impact of a cosmic ray (CR) background generated by supernova explosions from the first stars on star-forming metal-free gas in a minihalo at $z\sim25$.  
Starting from cosmological initial conditions, we use the smoothed particle hydrodynamics code GADGET-2 to follow gas collapsing under the influence of a CR background up to densities of $n=10^{12}\cc$, at which point we form sink particles. 
Using a suite of simulations with two sets of initial conditions and employing a range of CR background models, we follow each simulation for $5000\yr$ after the first sink forms.  
CRs both heat and ionise the gas, boosting $\htwo$ formation.  
Additional $\htwo$ enhances the cooling efficiency of the gas, allowing it to fulfil the Rees-Ostriker criterion sooner and expediting the collapse, such that each simulation reaches high densities at a different epoch. 
As it exits the loitering phase, the thermodynamic path of the collapsing gas converges to that seen in the absence of any CR background. 
By the time the gas approaches sink formation densities, the thermodynamic state of the gas is thus remarkably similar across all simulations. 
This leads to a robust characteristic mass that is largely independent of the CR background, of order $\sim$ a few $\times10\msun$ even as the CR background strength varies by five orders of magnitude.
\end{abstract}
\bibliographystyle{mnras}

\begin{keywords}
stars: formation --- stars: Population III --- cosmology: theory --- early Universe --- dark ages, first stars
\end{keywords}

\section{Introduction}
\label{intro}

Cosmic rays (CRs) have long been known to play an important role in the complex physics and chemistry of the interstellar medium (ISM) in local galaxies (\citealt{SpitzerTomasko1968,SpitzerScott1969,GlassgoldLanger1973,GoldsmithLanger1978,CravensDalgarno1978,MannheimSchlickeiser1994,Tielens2005}; recently reviewed by \citealt{StrongMoskalenkoPtuskin2007,GrenierBlackStrong2015}).  
They are an effective source of ionisation and heating in various environments, from preheating the primordial intergalactic medium  \citep[IGM;][]{SazonovSunyaev2015}, to driving outflows in the diffuse ISM \citep[e.g.,][]{Ensslinetal2007,Jubelgasetal2008,SalemBryan2014,Hanaszetal2013,Boothetal2013,SalemBryanHummels2014}, to providing an important (often dominant) source of heating and ionisation in deeply embedded gas clouds and protostellar discs \citep{Spitzer1978,DalgarnoYanLiu1999,IndrioloFieldsMcCall2009,PadovaniGalliGlassgold2009,GlassgoldGalliPadovani2012,PadovaniHennebelleGalli2013,Padovanietal2015}. 

CRs are particularly interesting in the context of Population III (Pop III) star formation, as they provide a continually replenished source of free electrons, enhancing the formation of molecular hydrogen---the only coolant available in primordial gas \citep{Abeletal1997,GalliPalla1998,BrommCoppiLarson2002}.  
This enhances the ability of the gas to cool, modifying the characteristic density and temperature at which runaway gravitational collapse sets in, and possibly the characteristic mass of the stars thus formed.   
The characteristic mass of Pop III is critical, as it largely controls the extent to which the first stars influence their environment, determining both their total luminosity and ionising radiation output \citep{Schaerer2002}, in addition to the details of their eventual demise \citep{Hegeretal2003,HegerWoosley2010,MaederMeynet2012}. 
As such, a thorough understanding of how the very first stars impact subsequent episodes of metal-free star formation---sometimes referred to as Pop III.1 and Pop III.2, respectively \citep{McKeeTan2008}---is crucial to developing a comprehensive picture of cosmic evolution.

In the absence of any feedback, pioneering numerical studies suggested that the very first stars were quite massive---on the order of $100\msun$ \citep[e.g.,][]{BrommCoppiLarson1999,BrommCoppiLarson2002,AbelBryanNorman2002,Yoshidaetal2003,BrommLarson2004,Yoshidaetal2006,OSheaNorman2007}. 
However, more recent simulations, aided by increased resolution, have found that significant fragmentation occurs during the star formation process \citep{StacyGreifBromm2010,Clarketal2011a,Clarketal2011b,Greifetal2011,Greifetal2012,StacyBromm2013,Hiranoetal2014,Hosokawaetal2015}, leading to the emerging consensus that the Pop III initial mass function (IMF) was somewhat top-heavy with a characteristic mass of $\sim$ a few $\times 10\msun$ \citep{Bromm2013}. 
While the contribution these stars make to chemical enrichment (\citealt{MadauFerraraRees2001,MoriFerraraMadau2002,BrommYoshidaHernquist2003,Hegeretal2003,UmedaNomoto2003,TornatoreFerraraSchneider2007,Greifetal2007,Greifetal2010,WiseAbel2008,Maioetal2011}; recently reviewed in \citealt{KarlssonBrommHawthorn2013}) and  reionisation \citep{Kitayamaetal2004,Sokasianetal2004,WhalenAbelNorman2004,AlvarezBrommShapiro2006,JohnsonGreifBromm2007,Robertsonetal2010} have been well studied, the consequences for Pop III stars forming in neighbouring minihaloes have been less thoroughly explored.  

As Pop III stars form in a predominantly neutral medium, the majority of their ionising output is absorbed, allowing only radiation less energetic than the Lyman-$\alpha$ transition to escape the immediate vicinity of the star-forming halo.  
While far-ultraviolet radiation in the Lyman-Werner (LW) bands ($11.2 - 13.6\ev$) lacks sufficient energy to interact with atomic hydrogen, it can still effectively dissociate molecular hydrogen.  
However, studies have found that the expected mean value of such radiation is far below the critical LW flux required to suppress $\htwo$ cooling \mbox{\citep{Dijkstraetal2008}}. 
At the high energy end, the neutral hydrogen cross section for X-rays and cosmic rays is small, allowing them to easily escape their host minihaloes. 
We recently investigated the impact of a cosmic X-ray background generated by high-mass X-ray binaries on primordial star formation \citep{Hummeletal2015}; here we focus on the impact of a CR background consisting of particles accelerated in supernova shock waves via the first-order Fermi process (see Section \ref{sec:context}).  

While the uncertainties involved in estimating the strength of the high-$z$  CR background are huge,  measurements of the $^6{\rm Li}$ abundance from metal-poor stars in the Galactic halo provide a useful constraint: the observed abundance of  $^6{\rm Li}$ is approximately 1000 times higher than predicted by big bang nucleosynthesis \citep{Asplundetal2006}, strong evidence for the existence of a CR spallation channel to provide a $^6{\rm Li}$ bedrock abundance prior to the bulk of star formation \citep{RollindeVangioniOlive2005,RollindeVangioniOlive2006}. 
Production of this first pervasive CR background by shock-acceleration in Pop III supernovae dovetails nicely with the upper limits this places on the CR energy density at high redshifts \citep{RollindeVangioniOlive2006}.

Early studies of the impact of CRs on primordial star formation focused on the production of ultra-high-energy CRs (UHECRs) by the decay of ultra-heavy X particles \citep{ShchekinovVasiliev2004,VasilievShchekinov2006,RipamontiMapelliFerrara2007}.  
With energies above the Greisen--Zatsepin--Kuzmin (GZK) cutoff \citep{Greisen1966,ZatsepinKuzmin1966}, UHECRs interact with the cosmic microwave background (CMB) to produce ionising photons, which in turn enhance the free electron fraction of the gas.  
Other work investigated the direct collisional ionisation of neutral hydrogen by SN shock-generated CRs, using one-zone models to determine the impact of a CR background on the chemical and thermal evolution of the gas in a minihalo \citep{StacyBromm2007,JascheCiardiEnsslin2007}.  
These studies found that the presence of a CR background enhanced molecular hydrogen formation in the minihalo, cooling the gas and lowering the Jeans mass, and by extension, the characteristic mass of the stars formed. 
We expand upon these one-zone models, using three-dimensional \textit{ab initio} cosmological simulations to investigate the impact of a CR background on Pop III stars forming in a minihalo.

This paper is organized as follows: In Section \ref{sec:context} we provide the cosmological context for this study, estimating the expected intensity of the CR background. Our numerical methodology is described in Section \ref{sec:methods}, while our results are presented in Section \ref{sec:results}.  
Finally, our conclusions are gathered in Section \ref{conclusions}. Throughout this paper we adopt a $\Lambda$CDM model of hierarchical structure formation, using the following cosmological parameters, consistent with the latest measurements from the Planck Collaboration \citep{PlanckParams2015}: $\Omega_{\Lambda} = 0.7$, $\Omega_{\rm m} = 0.3$, $\Omega_{\rm B} = 0.04$, and $H_0 = 70 \kms \Mpc^{-1}$.

\section{Cosmic Rays in the Early Universe}
\label{sec:context}
While there are several possible sources of CRs at high redshifts including primordial black holes, topological defects, supermassive particles, and structure formation shocks, the most likely source in the early Universe is supernova (SN) explosions \citep[e.g.,][]{GinzburgSyrovatskii1969,BiermannSigl2001,Stanev2004,Pfrommeretal2006}, wherein CRs are accelerated by the SN shock wave via the first-order Fermi process \citep[e.g.,][]{Bell1978}.  
In this scenario, high-energy particles diffuse back and forth across the shock front, increasing their energy by a small percentage each time, and resulting in a differential spectrum of CR number density per energy \citep{Longair1994} of the form
\begin{equation}
    \frac{{\rm d}n_{\rm \small CR}}{{\rm d}\epsilon} = \frac{n_{\rm norm}}{\epsilon_{\rm min}}
    \left( \frac{\epsilon}{\epsilon_{\rm min}} \right)^{-2},
\end{equation}
where $n_{\rm \small CR}$ is the CR number density, $\epsilon$ is the kinetic energy of the CR, $\epsilon_{\rm min}$ is the low-energy cutoff of the CR spectrum, and $n_{\rm norm}$ is a normalising density factor. 

The value of $\epsilon_{\rm min}$ is crucial, as the ionisation cross-section of non-relativistic CRs varies roughly as $\epsilon^{-1}$ for $\epsilon \gtrsim 10^5\ev$. 
Higher energy CRs thus travel much farther between interactions and are less able to effectively deposit their energy into the gas, such that the low-energy end of the CR spectrum provides the majority of the ionisation and heating.  
While estimates of the lowest possible energy CR protons could gain in a SN shock wave vary significantly, below $10^5\ev$ the velocity of the CR drops below the approximate orbital velocity of electrons in the ground state of atomic hydrogen, and the interaction cross-section diminishes rapidly \citep{Schlickeiser2002}. 

We therefore employ $\epsilon_{\rm min} = 10^6\ev$ as the lower bound for the CR background in our simulations.  
While CRs with energies below $10^6\ev$ may account for $\sim$$5-50$ percent of the CR energy budget, they also deposit a substantial fraction of their energy into the IGM \citep{SazonovSunyaev2015}. 
Consequently, they are unable to efficiently contribute to the build-up of a large-scale CR background. We thus neglect the contribution such CRs make to the ionisation and heating of primordial gas. 
However, this introduces only minor errors, as shown by \citet{StacyBromm2007}. Using one-zone models, they found that extending the CR spectrum to as low as $\epsilon_{\rm min} = 10^3\ev$ altered the resulting ionisation and heating rates by less than a factor of two.  
It should also be noted that the uncertainty in $\epsilon_{\rm min}$ is on par with that for the fraction of the SN energy going into CR production, $f_{\rm \small CR}$, which is expected to be $\sim$$10-20$ percent \citep{CaprioliSpitkovsky2014}.  
Here we assume $f_{\rm \small CR} = 0.1$.

The upper limit of the CR energy spectrum $\epsilon_{\rm max}$ depends on the strength of the ambient magnetic field in the local ISM, as the Fermi acceleration process is linearly dependent on the magnetic field through which the SN shock wave propagates. 
As the strength, generation mechanism, and distribution of magnetic fields in the early Universe remain highly uncertain \citep{DurrerNeronov2013}, we set $\epsilon_{\rm max} = 10^{15}\ev$, a typical maximum value from Fermi acceleration theory in SN shocks \citep[e.g.,][]{BlandfordEichler1987}.  
So long as $\epsilon_{\rm max}$ is below the GZK cutoff, $\epsilon_{\rm \small GZK}$, at that redshift, given by \citep{StacyBromm2007}
\begin{equation}
\epsilon_{\rm \small GZK}(z) = \frac{5\times10^{19}\ev}{1+z},
\end{equation}
the precise value of $\epsilon_{\rm max}$ is not crucial, since the great majority of the heating and ionisation is provided by CRs near $\epsilon_{\rm min}$.

Magnetic fields also play an important role in isotropising the CR background.  
We may estimate the effective mean free path over which a CR can freely propagate before being significantly impacted by the presence of a magnetic field $B$ using the Larmor radius $r_L$, where for a proton, 
\begin{equation}
r_L = \frac{\gamma \mh \beta c^2}{eB}.
\end{equation}
Here, $\gamma$ is the Lorentz factor of the CR; $\mh$, the proton mass; $e$, the proton charge; $c$, the speed of light, and $\beta c$ is the CR velocity. 
The CR background at redshift $z$ for protons with energy $\epsilon$ may then be considered fully isotropic if $r_L(\epsilon) \ll \Dhubble(z)$ where $\Dhubble(z)$ is the Hubble distance at that epoch. 
Recent observations have placed lower bounds on the modern-day intergalactic magnetic field strength ranging from $10^{-18}\,$G \citep{Dermeretal2011} to $3\times10^{-16}\,$G \citep{NeronovVovk2010}, while other estimates place the field strength as high as $10^{-15}\,$G \citep{AndoKusenko2010}.  
Simply accounting for flux freezing, the magnetic field at redshift $z$ is given by
\begin{equation}
B(z) = B_0 (1+z)^2,
\end{equation}
where $B_0$ is the magnetic field at the current epoch. 
The magnetic field strength in the IGM at $z=20$ then was likely in the range $4\times10^{-16}$ to $4\times10^{-13}\,$G.  
The magnetic field strength required to fully randomize the path of a $10^{15}\ev$ CR proton is $\sim$$10^{-14}\,$G; we may therefore assume the CR background is fully isotropic to the limit of our CR spectrum.  
The Larmor radius for a $10^6\ev$ CR proton in a $10^{-15}\,$G magnetic field, on the other hand, is $\sim$$10\pc$.  
While unable to freely propagate, under the assumption that magnetic fields at these redshifts are disordered on $\pc$ scales, such CRs will diffuse through the IGM with a diffusion coefficient $D = c r_{\rm L}$.  
Given that the typical physical distance $L$ between minihaloes at $z=20$---derived from the Press-Schechter (\citeyear{PressSchechter1974}) formalism---is $\sim$$5\kpc$, the diffusion time between minihaloes, $t_{\rm diff} = L^2/D$, is of order $10^{14}\s$, less than the Hubble time at this epoch.
$10^6\ev$ CRs are thus able to build up a locally isotropic background on scales larger than the typical distance between minihaloes.
On the scale of our simulation boxes (see Section \ref{setup}), CRs are thus able to build up an effectively uniform and isotropic background from $10^6$ to $10^{15}\ev$, an assumption we make throughout the remainder of our analysis.

Normalising the differential CR spectrum with the aforementioned limits $\epsilon_{\rm min}$ and $\epsilon_{\rm max}$ to the total energy density in cosmic rays, $\ucr$, results in a CR energy spectrum increasing over cosmic time in the following fashion:
 \begin{equation}
 \frac{{\rm d}n_{\rm \small CR}}{{\rm d}\epsilon}(z) = \frac{u_{\rm \small CR}(z)}{\epsilon_{\rm min}^2{\rm ln}\,\epsilon_{\rm max}{\large /}\epsilon_{\rm min}}  \left( \frac{\epsilon}{\epsilon_{\rm min}} \right)^{-2}.
 \end{equation}
Here we estimate $\ucr$ as follows:
\begin{equation}
u_{\rm \small CR}(z) = f_{\rm \small CR} E_{\rm \small SN}\, f_{\rm \small SN} \Psi_{*}(z)\, t_{\rm \small H}(z) (1+z)^3,
\end{equation}
where $f_{\rm \small CR}$ is the fraction of the SN explosion energy $E_{\rm \small SN}$ going into CR production, $f_{\rm \small SN}$ is the mass fraction of stars formed which die as SNe, and $\Psi_{*}(z)$ is the comoving star formation rate density (SFRD) as a function of redshift, which we assume to be constant over a Hubble time $t_{\rm \small H}$. 
The factor of $(1+z)^3$ accounts for the conversion from a comoving SFRD to a physical energy density. As in \citet{Hummeletal2015}, we base our estimate of $u_{\rm \small CR}(z)$ on the Pop III SFRD calculated by \citet{GreifBromm2006}, but see \citet{Campisietal2011} for a more recent calculation. 

The mass fraction of Pop III stars dying as SNe depends strongly on their IMF, and while the complex physical processes at play in gas collapsing from IGM to protostellar densities have so far prevented a definitive answer to this question, the emerging consensus is that the Pop III IMF was somewhat top-heavy with a characteristic mass of $\sim$ a few $\times 10\msun$ \citep{Bromm2013}.  
Given this, we assume one SN is produced for every $50\msun$ of stars formed and each SN-producing star dies quickly as a core-collapse explosion with $E_{\rm \small SN} = 10^{51}\erg$, 10 percent of which goes into CR production.  

\begin{figure}
\begin{center}
\includegraphics[width=1\columnwidth]{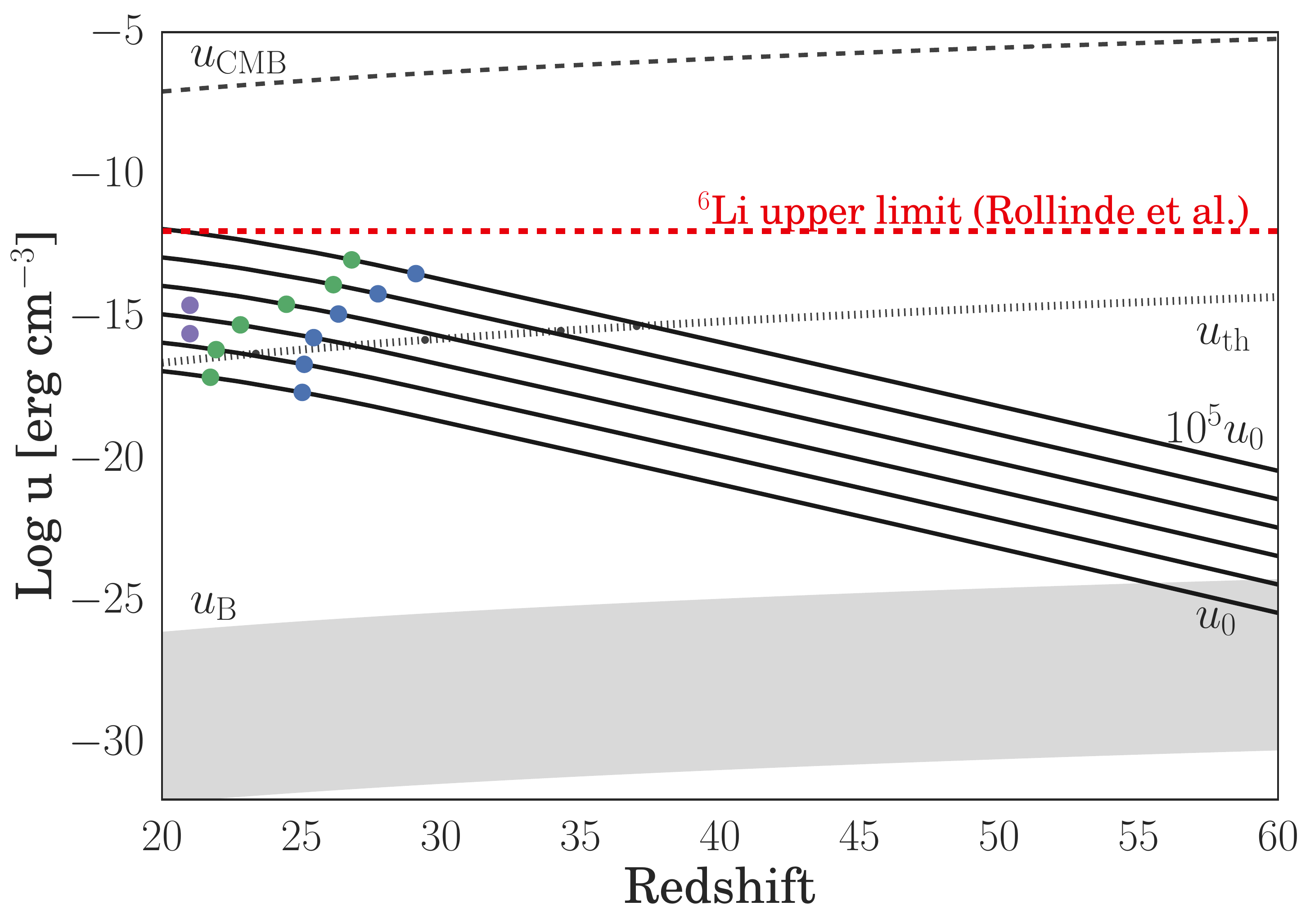}
\caption{\label{fig:ucr}
Cosmic ray energy density $\ucr$ as a function of redshift.  
Our models, from $u_0$ to $10^5u_0$ are indicated by solid black lines, while the energy density in the CMB and the thermal energy density in the IGM are denoted by dashed and dotted black lines, respectively. 
The grey region shows the range of possible values for the magnetic field energy density in the IGM, and the dashed red line marks the most conservative upper limit on the CR energy density from \citet{RollindeVangioniOlive2006}, as derived from the $^6$Li abundance. 
Blue points indicate the redshift at which Halo 1 collapses to high densities for a given $\ucr$; green points, the redshift at which Halo 2 collapses.  
Finally, purple points denote the energy densities employed by \citet{StacyBromm2007} to study the impact of a CR background using one-zone models calculated at $z=21$.%
}
\end{center}
\end{figure}

This fiducial estimate for $\ucrz$---henceforth referred to as model $u_0$---is shown in Figure \ref{fig:ucr}, where we compare the energy density in CRs to the gas thermal energy density in the IGM, the range of estimates for the IGM magnetic field energy density, and the energy density of the cosmic microwave background. 
We also mark the most conservative upper limits placed on $\ucr$ by \citet{RollindeVangioniOlive2006}, calculated using the observed overabundance of $^6$Li compared to big bang nucleosynthesis. 
Given the huge uncertainties inherent in estimating the strength of the high-$z$ CR background, we also consider five additional models with 10, $10^2$, $10^3$, $10^4$, and $10^5$ times the energy density of model $u_0$, as shown in Figure \ref{fig:ucr}.  
This allows us to bracket the plausible range of energy densities, from values comparable to the IGM thermal energy density to just below the upper limits from $^6$Li abundance measurements.

\section{Numerical Methodology}
\label{sec:methods}
Using the well-tested $N$-body smoothed particle hydrodynamics (SPH) code GADGET-2 \citep{Springel2005}, we perform our simulations using the same chemistry network and sink particle method as described in \citet{Hummeletal2015}; the relevant details are summarized below.  
In addition to using the same initial conditions as in \citet{Hummeletal2015} to allow for a comparison of the impact of X-rays versus cosmic rays on the primordial gas, we perform a second suite of simulations using new initial conditions.  
The minihalo in these simulations collapses at a lower redshift; this allows us to ensure our results are not being influenced by the CMB temperature floor.

\subsection{Initial Setup}
\label{setup}
The first set of simulations (henceforth referred to as Halo 1) use the same initial conditions as \citet{Hummeletal2015} and \citet{StacyGreifBromm2010}, and are initialised at $z=100$ in a 140 comoving kpc box with periodic boundary conditions. 
To accelerate structure formation within the simulation box an artificially enhanced normalisation of the power spectrum, $\sigma_8 = 1.4$, was used, but the simulations are otherwise initialised in accordance with a $\Lambda$CDM model of hierarchical structure formation. 
For a discussion of the validity of this choice, see \citet{StacyGreifBromm2010}. 
High resolution in these simulations is achieved using a standard hierarchical zoom-in technique, with nested levels of refinement at 40, 30, and 20 kpc (comoving).  
Using this technique, the highest resolution SPH particles have a mass $m_{\rm SPH} = 0.015\msun$, yielding a minimum mass resolution $M_{\rm res} \simeq 1.5 N_{\rm neigh} m_{\rm SPH} \lesssim 1\msun$ for the simulation.  
Here $N_{\rm neigh} = 32$ is the number of particles used in the SPH smoothing kernel \citep{BateBurkert1997}.

The initial conditions for the second set of simulations (henceforth referred to as Halo 2) are derived from the simulations described in \citet{StacyBromm2013}.  
These simulations were initialised at $z=100$ in a 1.4 comoving Mpc box in accordance with the same $\Lambda$CDM structure formation model as Halo 1, but with $\sigma_8 = 0.9$ rather than the artificially enhanced $\sigma_8 = 1.4$ used in Halo 1. 
The use of `marker sinks'  for regions reaching densities beyond $n=10^3\,{\rm cm}^{-3}$ rather than following the gas to higher densities allows for the efficient identification of the first 10 minihaloes to form within the box; we select the final minihalo to form---i.e., Minihalo 10 from \citet{StacyBromm2013}---for further study.  
After identifying the location of the final minihalo, the cosmological box is re-initialised at $z=100$ using a standard zoom-in technique with two nested levels of refinement used to improve resolution surrounding the selected minihalo.  
Each `parent' particle within the most refined region is split into 64 `child' particles, with a minimum mass of $m_{\rm gas}=1.85\msun$ and $m_{\rm \small DM}=12\msun$. 
These refined simulations are then run until the gas approaches the typical onset of runaway collapse at $n=10^4\cc$.

\subsubsection{Halo 2 Cut-out and Refinement}
\label{cutout}

Once the simulation reaches $10^4\cc$ all particles beyond 10 comoving kpc from the centre of the collapsing minihalo are removed to conserve computational resources. 
To achieve a maximum resolution comparable to Halo 1, all particles within the 10 comoving kpc volume are split into two child particles placed randomly within the smoothing kernel of the parent particle.  
This particle-splitting is repeated twice more; all particles within 8 comoving kpc are replaced by 8 child particles; within 6 comoving kpc, each of these child particles is split into an additional 8 for a total of 128 child particles in the most refined region.  
At each step, the mass of the parent particle is evenly divided among the child particles and the smoothing length is set to $h N_{\rm new}^{-1/3}$, where $h$ is the smoothing length of the parent and $N_{\rm new}$ is the number of child particles created.  
All particles inherit the same entropy, velocity, and chemical abundances as their parent, ensuring conservation of mass, internal energy, and momentum.

While our cut-out technique leads to a rarefaction wave propagating inward from the suddenly introduced vacuum boundary condition, the average sound speed at the edge of the most refined region is $\sim$$1\,{\rm km}\,{\rm s}^{-1}$.  
As a result the wave only travels $\sim$$0.5\,$pc over the remaining 400,000$\,$yr of the simulation, a negligible fraction of the $\sim$350$\,$pc physical box size.

\subsection{Chemistry and Thermodynamics}
\label{chemistry}
 We employ the chemistry network described in detail by \citet{Greifetal2009b}, which follows the abundance evolution of $\h$, $\hplus$, $\hminus$, $\htwo$, $\htwo^+$, $\he$, $\heplus$, $\he^{++}$, $\deut$, $\deut^+$, $\hd$ and e$^-$. 
 All relevant cooling mechanisms are accounted for, including $\h$ and $\he$ collisional excitation and ionisation, recombination, bremsstrahlung and inverse Compton scattering. 
 
 In order to properly model the chemical evolution at high densities, $\htwo$ cooling induced by collisions with $\h$ and $\he$ atoms and other $\htwo$ molecules is also included.  
 Three-body reactions involving $\htwo$ become important above $n \gtrsim 10^8\cc$; we employ the intermediate rate from \citet{PallaSalpeterStahler1983}, but see \citet{Turketal2011} for a discussion of the uncertainty of these rates. 
 In addition, the efficiency of $\htwo$ cooling is reduced above $\sim$$10^9\cc$ as the ro-vibrational lines of $\htwo$ become optically thick above this density.  
 To account for this we employ the Sobolev approximation together with an escape probability formalism (see \citealt{Yoshidaetal2006, Greifetal2011} for details). 

\subsection{Cosmic Ray Ionisation and Heating}
\label{CRchem}

Each time a CR proton ionises a hydrogen atom, an electron with average energy $\langle E \rangle = 35\ev$ is produced \citep{SpitzerTomasko1968}.  
Including the ionization energy of $13.6\ev$, the CR proton loses approximately $50\ev$ per scattering. 
This necessarily places a limit on how many scatterings a CR proton can undergo before losing all its energy to ionisation, as well as limiting the distance it may travel.  
This distance may be described by a penetration depth 
\begin{equation}
    D_p(n, \epsilon) \approx \frac{\beta c \epsilon} {-({\rm d}\epsilon / {\rm d}t)_{\rm ion}}
\end{equation}
where \citep{Schlickeiser2002}
\begin{equation}
    - \left( \frac{{\rm d}\epsilon} {{\rm d}t} \right)_{\rm ion}(n, \epsilon)
    = 1.82\times10^{-7}\,{\rm \small eV\,s}^{-1} n_{\h} f(\epsilon),
\end{equation}
\begin{equation}    
    f(\epsilon) = (1 + 0.0185 \,{\rm ln}\beta )\, \frac{2 \beta^2}{\beta_0^3 + 2 \beta^3},
\end{equation}
and 
\begin{equation}
    \beta =  \sqrt{1 - \left( \frac{\epsilon}{m_{\rm \tiny H}c^2}+1 \right)^{-2}}.
\end{equation}
Here, $({\rm d}\epsilon / {\rm d}t)_{\rm ion}$ is the rate at which a CR proton loses energy to ionisation. 
$\beta_0$ is the cutoff below which the interaction between CRs and the gas decreases sharply; we use $\beta_0=0.01$, appropriate for CRs travelling through a neutral IGM \citep{StacyBromm2007}.
As $D_p(n, \epsilon)$ is the mean free path of CRs of energy $\epsilon$ travelling through a gas with number density $n$, we may define an effective cross-section $\sigma_{CR}(n,\epsilon)$ for the interaction
\begin{equation}
\sigma_{CR}(n,\epsilon) = \frac{1}{n D_p(n, \epsilon)}.
\end{equation}

As the CR penetration depth is $\gg$ than the box size everywhere except approaching the centre of the star-forming minihalo, we may estimate the column density $N$---and thus, the CR attenuation along a given line of sight---using the same technique described in \citet{Hummeletal2015}. 
The gas column density approaching the centre of the accretion disk varies by roughly an order of magnitude between the polar ($N_{\rm \small pole}$) and equatorial ( $N_{\rm \small equator}$) directions, with the column density  along these lines of sight well fit by
\begin{equation}
{\rm log}_{10}(N_{\rm \small pole}) = 0.5323\, {\rm log_{10}}(n) + 19.64
\end{equation}
and
\begin{equation}
{\rm log}_{10}(N_{\rm \small equator}) = 0.6262\, {\rm log_{10}}(n) + 19.57, 
\end{equation}
respectively. We assume every line of sight within 45 degrees of the pole experiences column density $N_{\rm \small pole}$ while every other line of sight experiences $N_{\rm \small equator}$ \citep{Hosokawaetal2011}, allowing us to calculate an effective optical depth such that
\begin{equation}
e^{-\tau_{\rm \small CR}} = \frac{2 \Omega_{\rm \small pole}}{4\pi} e^{-\sigma_{\rm \small CR} N_{\rm \small pole}} + \frac{4\pi - 2 \Omega_{\rm \small pole}}{4\pi} e^{-\sigma_{\rm \small CR} N_{\rm \small eq}},
\end{equation}
where
\begin{equation}
\Omega_{\rm \small pole} = \int_0^{2\pi}{\rm d}\phi \int_0^{\pi/4}{\rm sin}\theta \,{\rm d}\theta = 1.84\,{\rm sr}.
\end{equation}

\begin{figure}
\begin{center}
\includegraphics[width=\columnwidth]{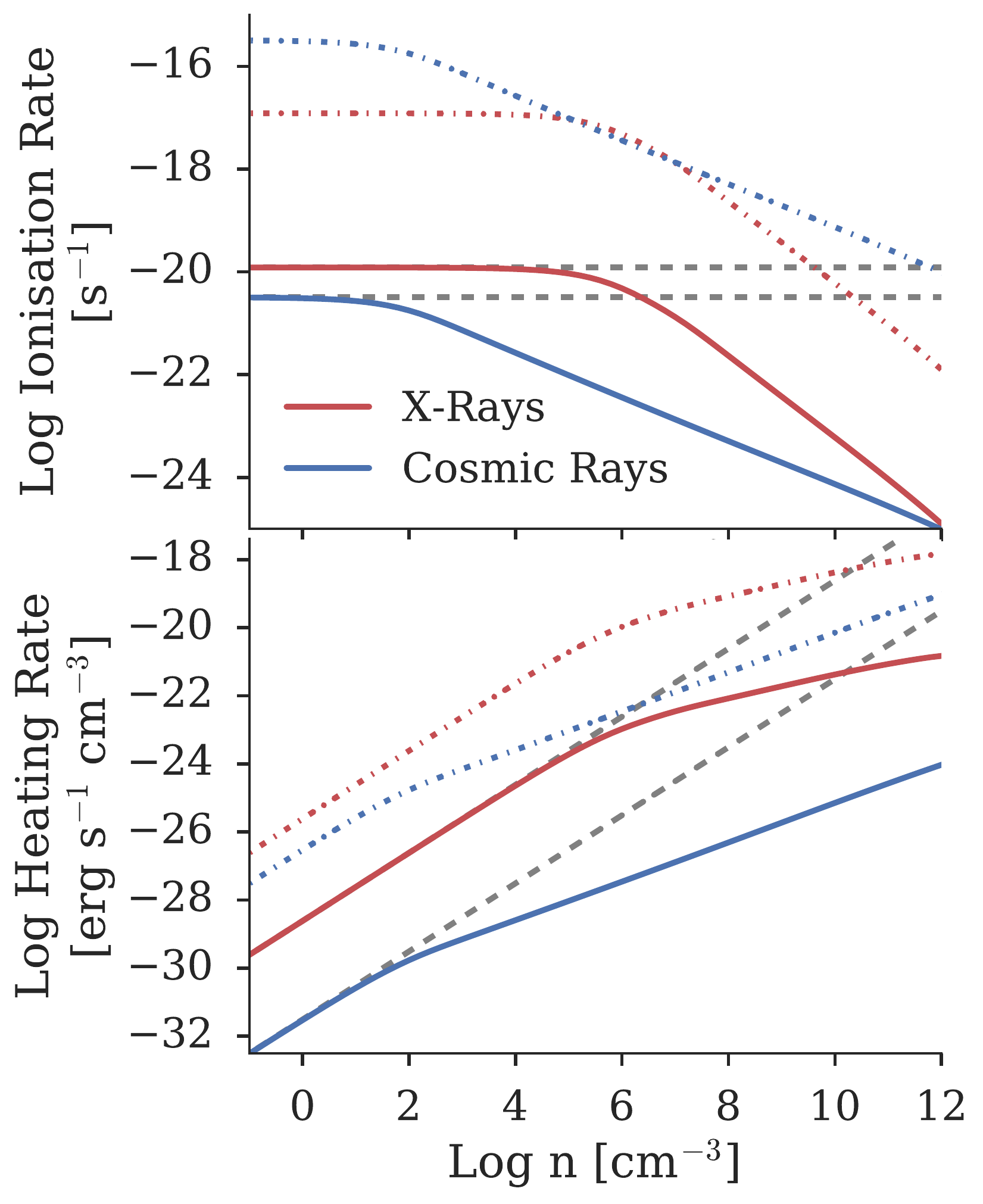}
\caption{\label{fig:khrates}
Ionisation and heating rates as a function of total gas number density for both a cosmic ray and an X-ray background at $z=25$. 
In both panels, the solid blue lines denote the rate for our fiducial CR background, $u_0$. The solid red lines show the fiducial X-ray rates from \citet{Hummeletal2015}, while the dashed grey lines demonstrate the expected rates in the absence of gas self-shielding.  
The dash-dotted blue (red) lines mark the highest-intensity CR (X-ray) heating and ionisation rates investigated. 
For a given energy density, CRs are more effective at ionising and heating the gas; vertical placement on this chart is simply a matter of normalisation, stemming from the assumptions outlined in Section \ref{sec:context}. 
Note that X-rays cause significantly more heating per ionisation event than CRs, but are strongly attenuated at high densities. CRs are significantly more penetrative in comparison.%
} 
\end{center}
\end{figure}

\begin{figure*}
\begin{center}
\includegraphics[width=.8\textwidth]{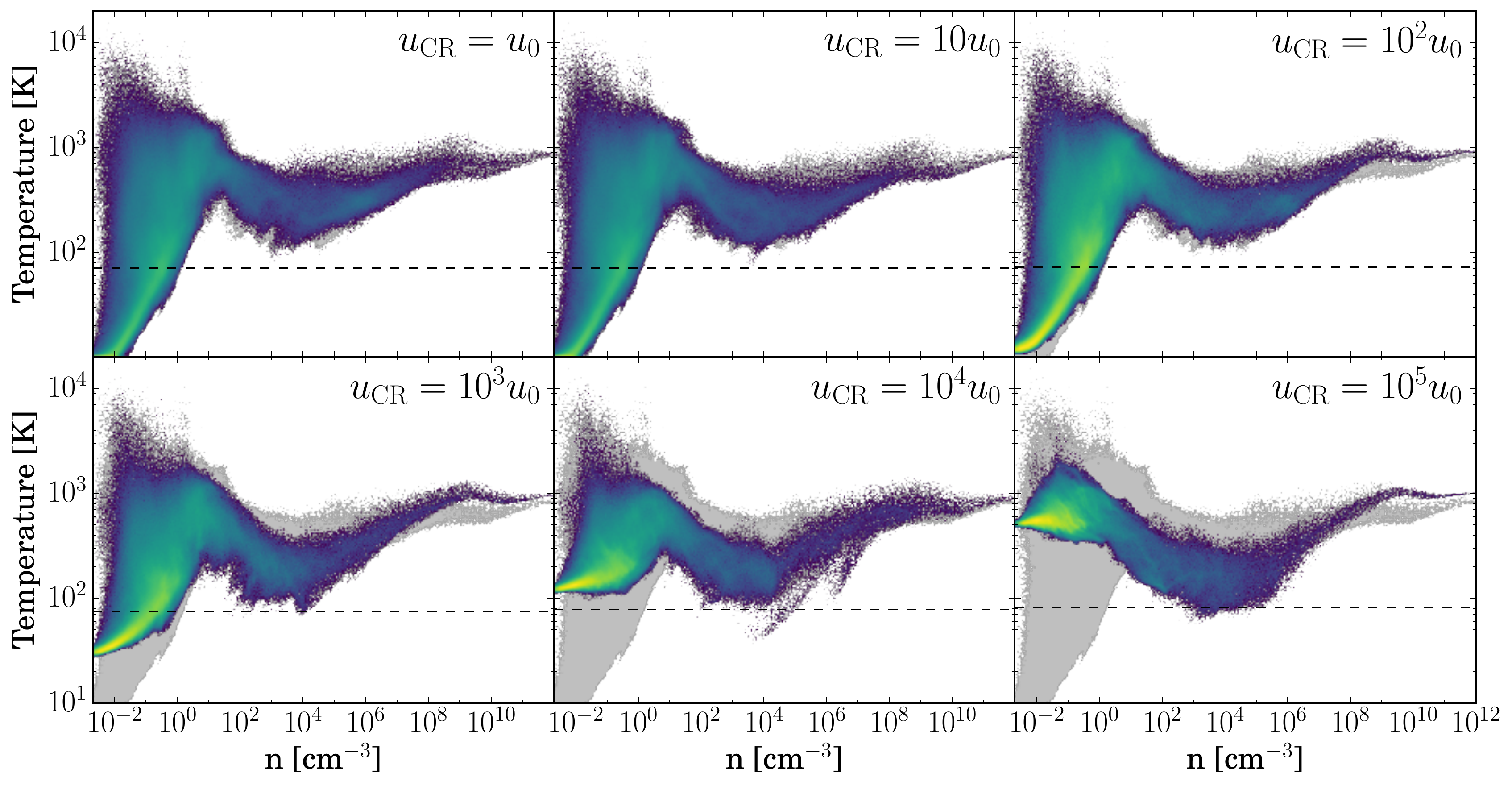}
\includegraphics[width=.8\textwidth]{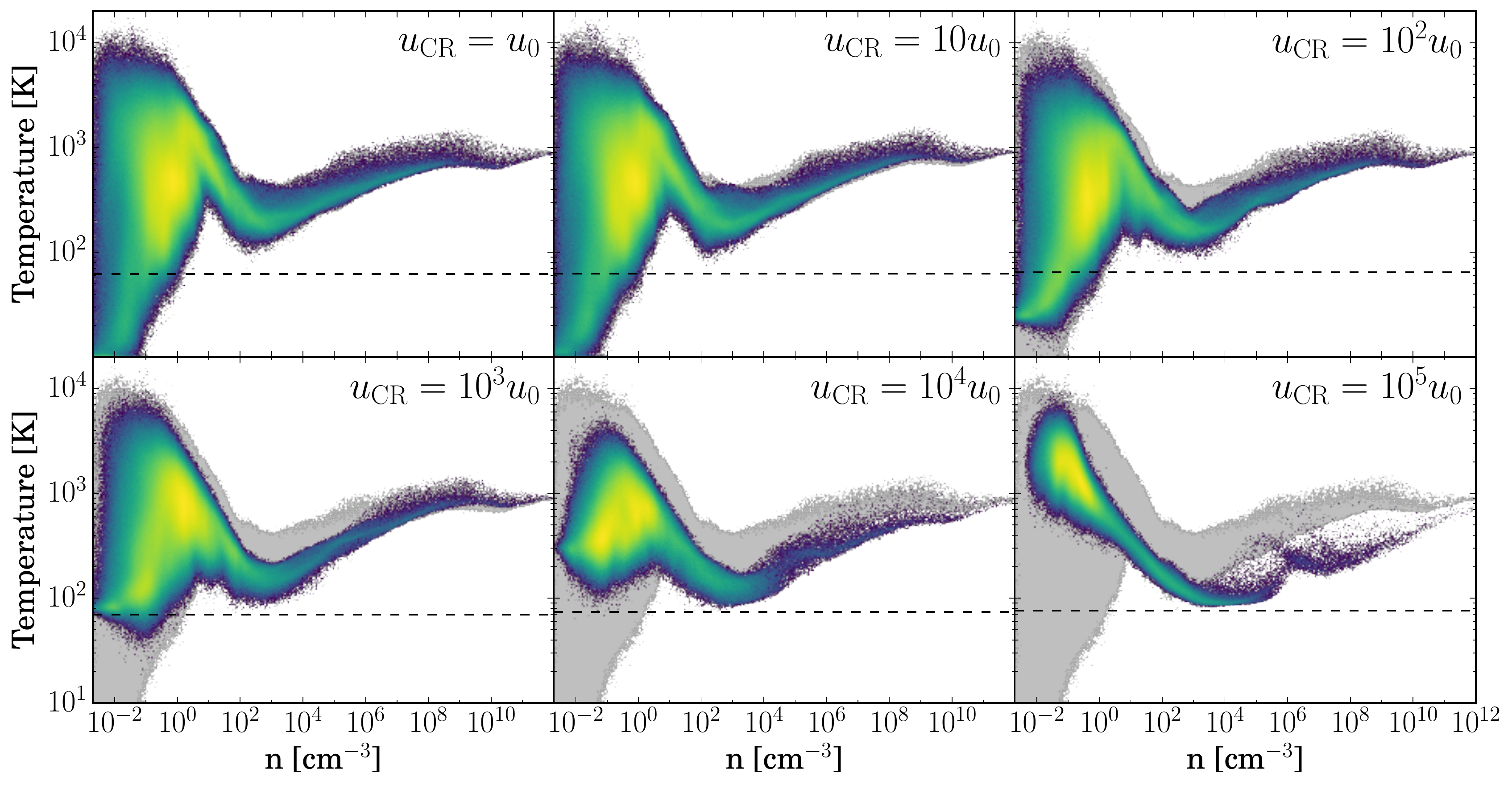}
\caption{\label{fig:temp}
Mass-weighted temperature-density distribution of the gas collapsing into Halo 1 (top) and Halo 2 (bottom), shown just prior to sink formation. 
Each panel shows the behaviour of gas collapsing under a CR background of a given strength. 
In each case, the behaviour of the gas in the $\ucr=0$ case is reproduced for comparison (light grey); dashed lines denote the CMB temperature when collapse occurs. Gas at low densities gets progressively hotter and gas in the loitering phase gets progressively cooler with increasing $\ucr$. 
Note that the gas invariably proceeds to collapse, and converges toward a similar thermodynamic path as it approaches sink formation densities, regardless of the CR background strength.%
}
\end{center}
\end{figure*}

Accounting for this attenuation of the CR background and following the treatment of \citet{StacyBromm2007} and \citet{InayoshiOmukai2011}, the cosmic ray heating rate $\Gamma_{\rm \small CR}$ and ionisation rate $k_{\rm \small CR}$ are given by
\begin{equation}
\Gamma_{\rm \small CR} = 
    \frac{ E_{\rm \small heat}}{50\,{\rm \small eV}} 
    \int_{\epsilon_{\rm min}}^{\epsilon_{\rm max}} 
    \left| \left( \frac{{\rm d}\epsilon} {{\rm d}t} \right)_{\rm ion} \right|
    \frac{dn_{\rm \tiny CR}}{d\epsilon} e^{-\tau_{\rm \small CR}} d\epsilon,
\end{equation}
and 
\begin{equation}
k_{\rm \small CR} = \frac{\Gamma_{\rm \small CR}}{ n_{\rm \small H} E_{\rm \small heat}},
\end{equation}
where $\epsilon_{\rm min} = 10^6\ev$, $\epsilon_{\rm max}= 10^{15}\ev$, $n_{\rm \small H}$ is the number density of hydrogen, and $E_{\rm \small heat}$ is the energy deposited as heat per interaction \citep{Schlickeiser2002}.
While CRs lose about $50\ev$ per interaction, only about $6\ev$ of that goes towards heating in a neutral medium \citep{SpitzerScott1969,ShullvanSteenberg1985}; we set $E_{\rm \small heat}$ accordingly.

Here we assume the incident CR background is composed solely of protons, and all interactions occur with hydrogen only.  
While this neglects the slight difference in average CR energy loss per interaction for hydrogen ($36\ev$; \citealt{BakkerSegre1951}) as compared to helium ($40\ev$; \citealt{WeissBernstein1956}), the resulting error in the employed heating and ionisation rates is sufficiently small for our purposes (see \citealt{JascheCiardiEnsslin2007} for a more rigorous treatment). 

The resulting heating and ionisation rates for both the $\ucr=u_0$ and $10^5\,u_0$ cases are shown in Figure \ref{fig:khrates}, along with the expected rates in the absence of attenuation.
Also shown are the highest and lowest X-ray heating and ionisation rates considered in \citet{Hummeletal2015}. 
Compared to X-rays, CRs produce significantly less heating per ionisation event and penetrate to much higher densities before being attenuated; the X-ray heating and ionisation rates fall off much faster as the gas becomes optically thick.
\subsection{Sink Particles}
\label{sinkParticles}
Our sink particle method is described in \citet{StacyGreifBromm2010}. 
When a gas particle exceeds $n_{\rm max} = 10^{12}\cc$, it and all non-rotationally-supported particles within the accretion radius $r_{\rm acc}$ are replaced by a single sink particle.  
We set $r_{\rm acc}$ equal to the resolution length of the simulation: $r_{\rm acc} = L_{\rm res} \simeq 50\au$, where 
\begin{equation}
L_{\rm res} \simeq 0.5 \left( \frac{M_{\rm res}}{\rho_{\rm max}} \right)^{1/3},
\end{equation}
and $\rho_{\rm max} = n_{\rm max} m_{\rm H}$.  
Upon creation, the sink immediately accretes the majority of the particles within its smoothing kernel, resulting in an initial mass for the sink particle $M_{\rm sink}$ close to $M_{\rm res} \simeq 1\msun$.  
Following its creation, the density, temperature, and chemical abundances of the sink particle are no longer updated.  
The sink's density and temperature are held constant at $10^{12}\cc$ and $650\kelvin$, respectively; the pressure of the sink is set accordingly. 
Assigning a temperature and pressure to the sink particle allows it to behave as an SPH particle, thus avoiding the creation of an unphysical pressure vacuum which would artificially enhance the accretion rate onto the sink \citep[see][]{BrommCoppiLarson2002, MartelEvansShapiro2006}. 
Once the sink is formed, additional particles (including smaller sinks) are accreted as they approach within $r_{\rm acc}$ of that sink particle, and the position and momentum of the sink particle is set to the mass-weighted average of the pre-existing sink and the accreted particle.

\begin{figure}
\begin{center}
\includegraphics[width=1\columnwidth]{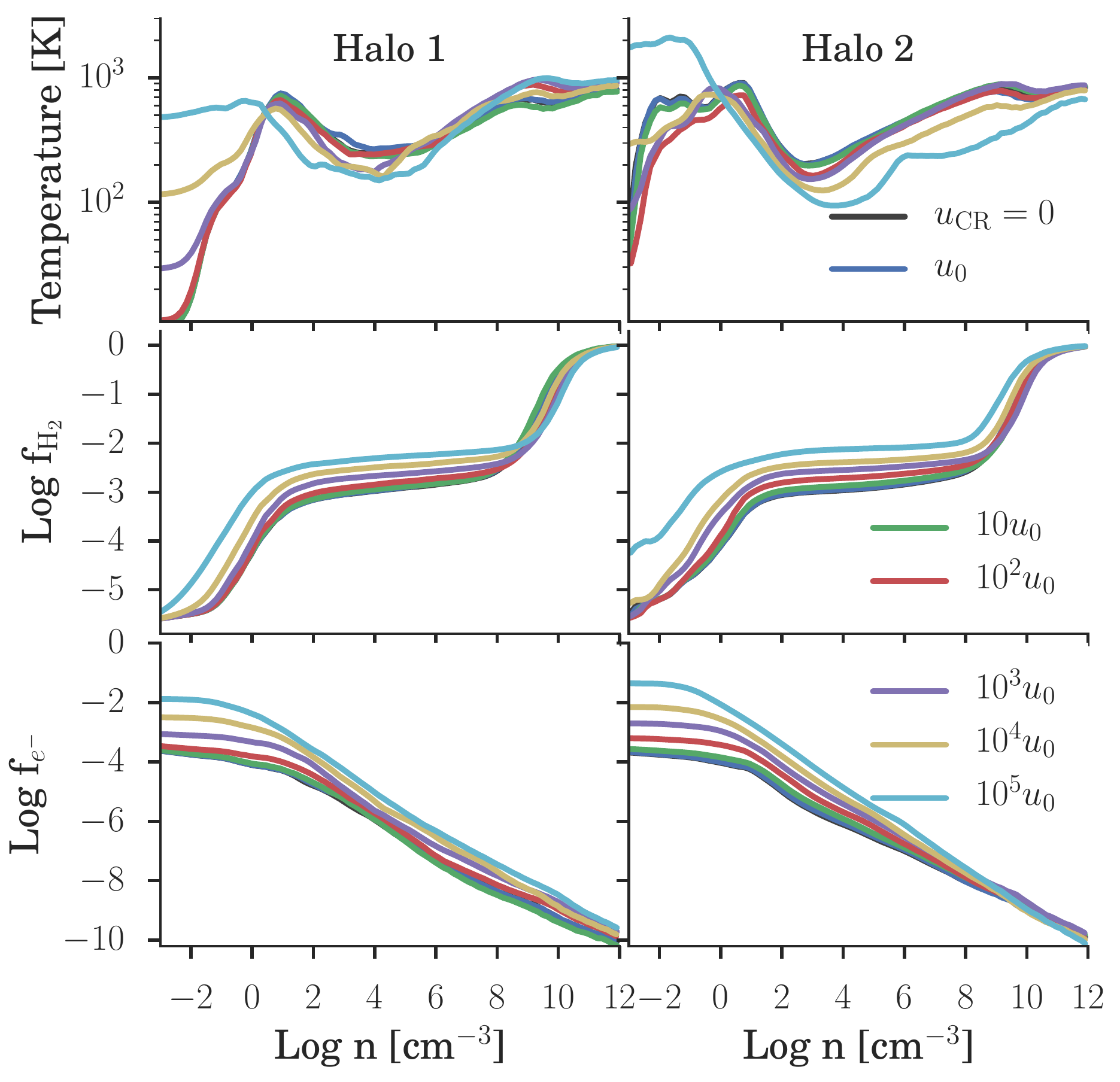}
\caption{\label{fig:efrac}
From top to bottom, density-binned average gas temperature, $\htwo$ fraction, and free electron fraction for both Halo 1 (left) and Halo 2  (right). 
As $\ucr$ increases, the additional ionisation increases f$_{e^-}$, which in turn elevates f$_{\htwo}$. 
Combined with the additional heating provided by the CR background, this allows $\htwo$ cooling to activate at slightly lower densities and gas in the loitering phase to reach slightly lower temperatures, as seen in the top panel.%
}
\end{center}
\end{figure}

\begin{figure}
\begin{center}
\includegraphics[width=.95\columnwidth]{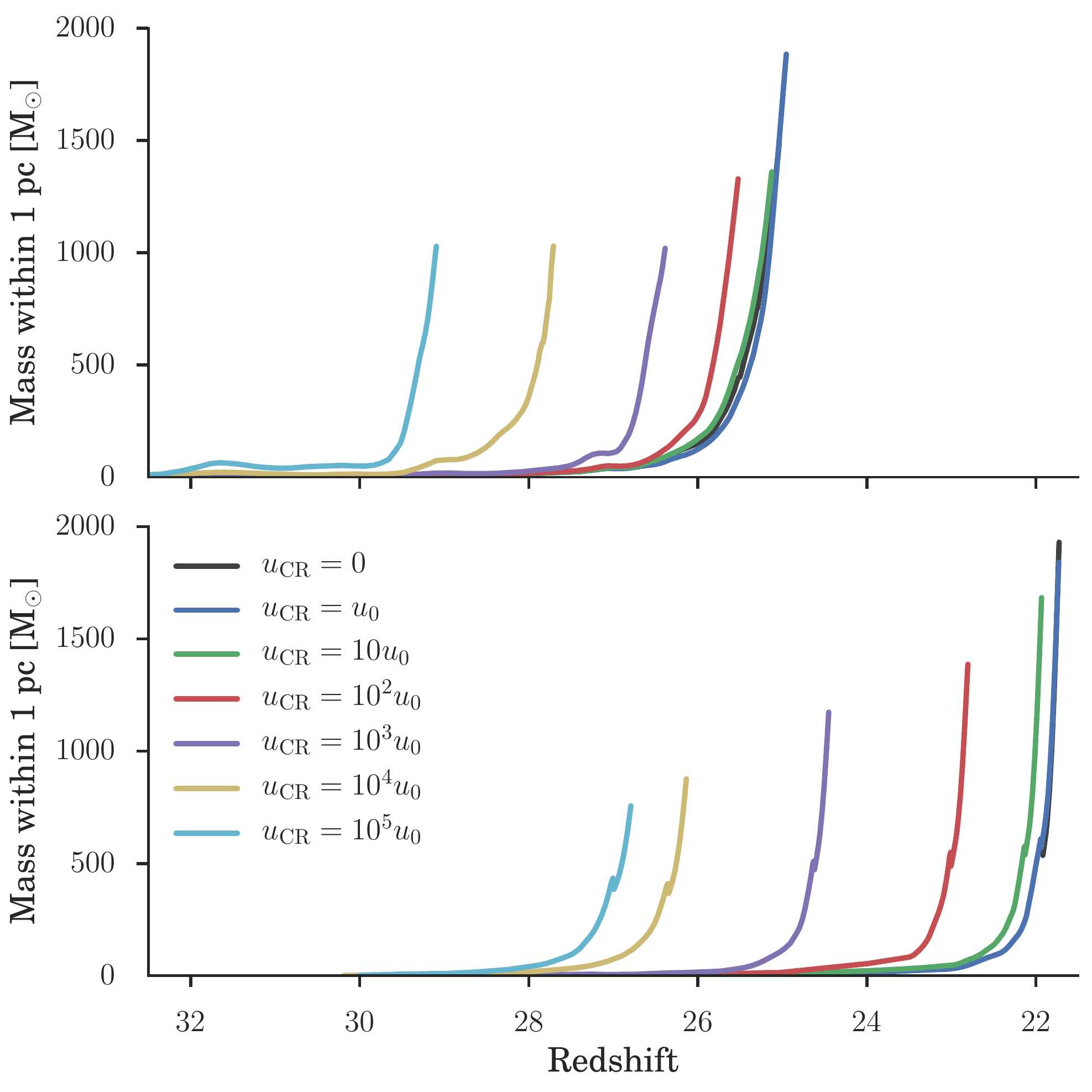}
\caption{\label{fig:collapse}
Total gas mass within $1\pc$ of the highest density point in each simulation over cosmic time.
Top: Halo 1; bottom: Halo 2.
As $\ucr$ increases, the additional heating and ionisation allows the gas to fulfil the Rees-Ostriker criterion ($t_{\rm cool} \lesssim t_{\rm ff}$) sooner, expediting collapse to high densities.  
The sooner runaway collapse sets in, the less time gas has to accumulate in the centre of the minihalo, resulting in a lower total mass within $1\pc$ at simulation's end.
The discontinuity seen in the growth of Halo 2 is a numerical artefact arising from the cut-out and refinement process.%
}
\end{center}
\end{figure}

\section{Results}
\label{sec:results}
\subsection{Initial Collapse}
\subsubsection{CR-enhanced H$_2$ Cooling}
\label{sec:initial_collapse}

The gas in both Halo 1 and Halo 2 invariably collapses to high densities, even as we vary the CR background strength by five orders of magnitude. 
As seen in Figure \ref{fig:temp}, the collapse occurs in accordance with the standard picture of Pop III star formation \citep[e.g.,][]{Greifetal2012,StacyBromm2013,Hiranoetal2014,Hosokawaetal2015}, albeit slightly modified by the presence of a CR background.  
In each case, the gas heats adiabatically as it collapses until reaching $\sim10^3\kelvin$, at which point $\htwo$ cooling is activated.  
This allows the gas to cool to $\sim200\kelvin$, whereupon it enters a `loitering' phase \citep{BrommCoppiLarson2002}, increasing in density quasi-hydrostatically until sufficient mass accumulates to trigger the Jeans instability, typically between $10^4\cc$ and $10^6\cc$. 
Runaway collapse then proceeds until three-body reactions become important at $n\sim10^8\cc$; this turns the gas fully molecular by $n\sim10^{12}\cc$, at which point we insert sink particles.

As demonstrated in Figure \ref{fig:efrac}, increasing $\ucr$ both heats the gas and boosts the resulting ionisation fraction.
This in turn increases the efficiency of $\htwo$ formation, enhancing cooling and allowing gas in the loitering phase to reach progressively lower temperatures. 
In fact, the additional cooling is sufficient to cool the gas all the way to the CMB floor when $\ucr \gtrsim 10^4\,u_0$. 
However as the gas exits the loitering phase and proceeds to higher densities, the CR optical depth increases significantly. 
In concert with the onset of three-body processes this renders CR ionisation and heating ineffective at high densities. 
As a result, the thermal behaviour of the gas increasingly resembles that seen in the $\ucr=0$ case as the collapse proceeds. 
To wit, by the time we form sink particles at  $10^{12}\cc$ the thermodynamic state of the gas is remarkably similar, even as we vary the CR background strength by five orders of magnitude.
This convergence under a wide range of environmental conditions is similar to that seen in the presence of both an X-ray background \citep{Hummeletal2015} and DM--baryon streaming \citep{StacyBrommLoeb2011a,Greifetal2011b}.

\begin{figure*}
\begin{center}
\includegraphics[width=.95\textwidth]{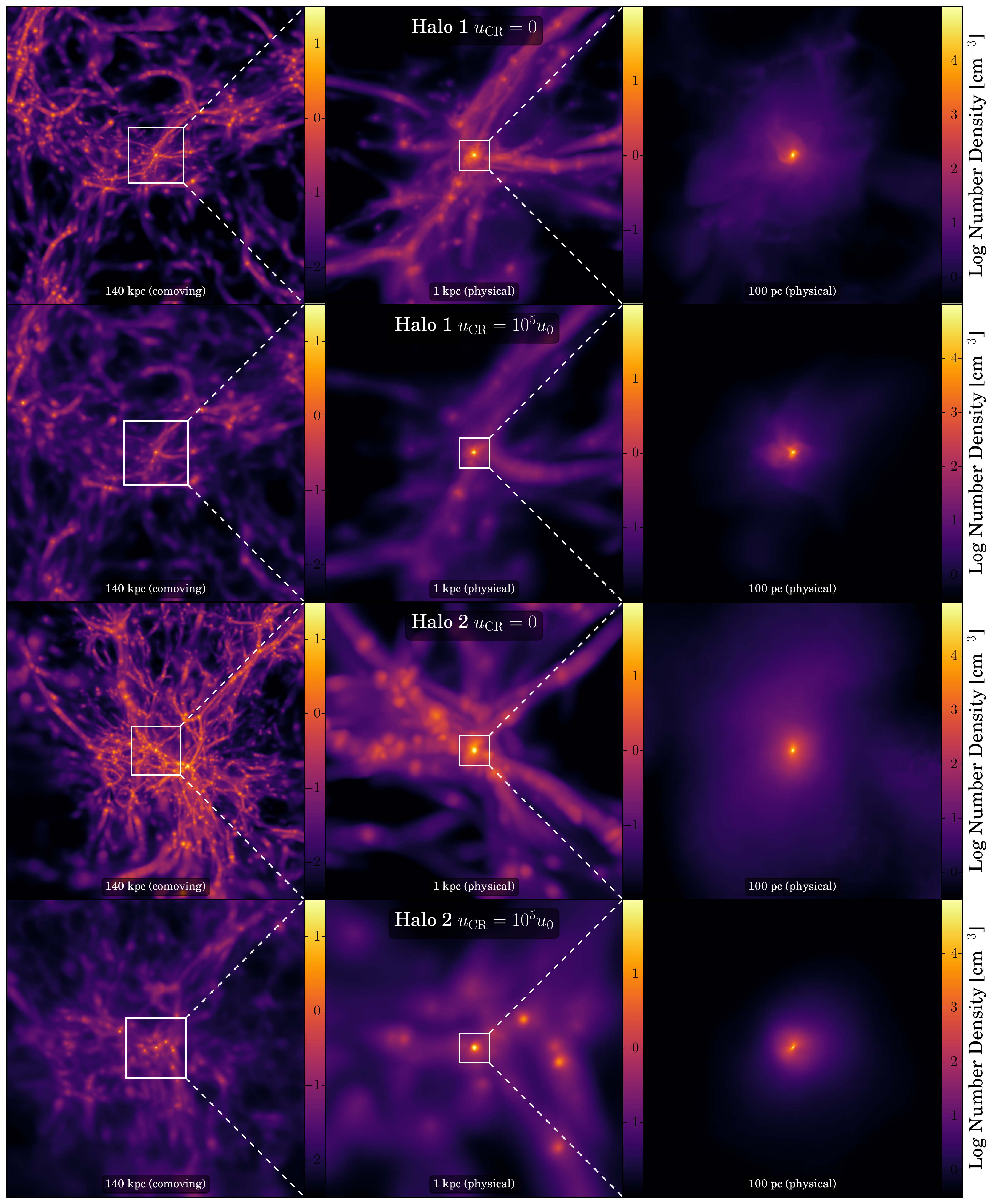}
\caption{\label{fig:structure}
Simulation structure for both Halo 1 (top) and Halo 2 (bottom) as seen in the final output, $5000\yr$ after the first sink forms.  
Shown is a density projection of the minihalo environment on progressively smaller scales for both the $\ucr=0$ and $\ucr=10^5\,u_0$ simulations, as labelled.  
White boxes indicate the region depicted on the next smaller scale.  
From left to right: filamentary structure of the cosmic web near the minihalo formation site; minihalo formation at the intersection of several filaments; morphology within the $\sim$$100\pc$ virial radius of the minihalo. 
The density scale for each level is shown just to the right -- note that the scaling changes from panel to panel.
Note how the gas in the $\ucr=10^5\,u_0$ simulations appears smoothed out compared to the $\ucr=0$ simulations.  
Not only do CRs heat and smooth the filamentary gas, the expedited collapse induced by CR ionisation leaves less time for low-density structure to form, since the collapse is shifted to an earlier epoch.  
As a result, the dynamical environment in which the sink particles form varies significantly between simulations.%
}
\end{center}
\end{figure*}

\begin{figure*}
\begin{center}
\includegraphics[width=.49\textwidth]{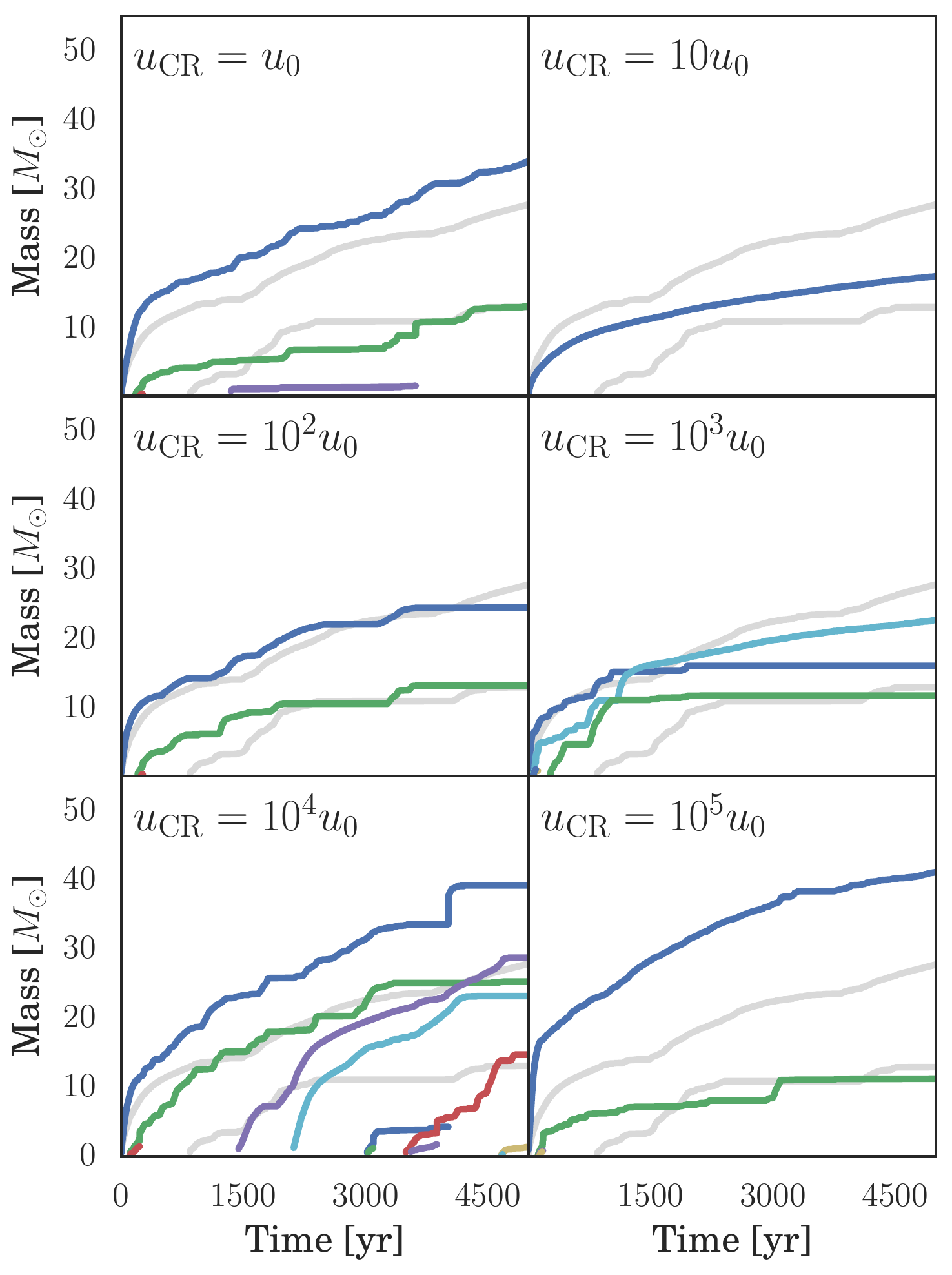}
\includegraphics[width=.49\textwidth]{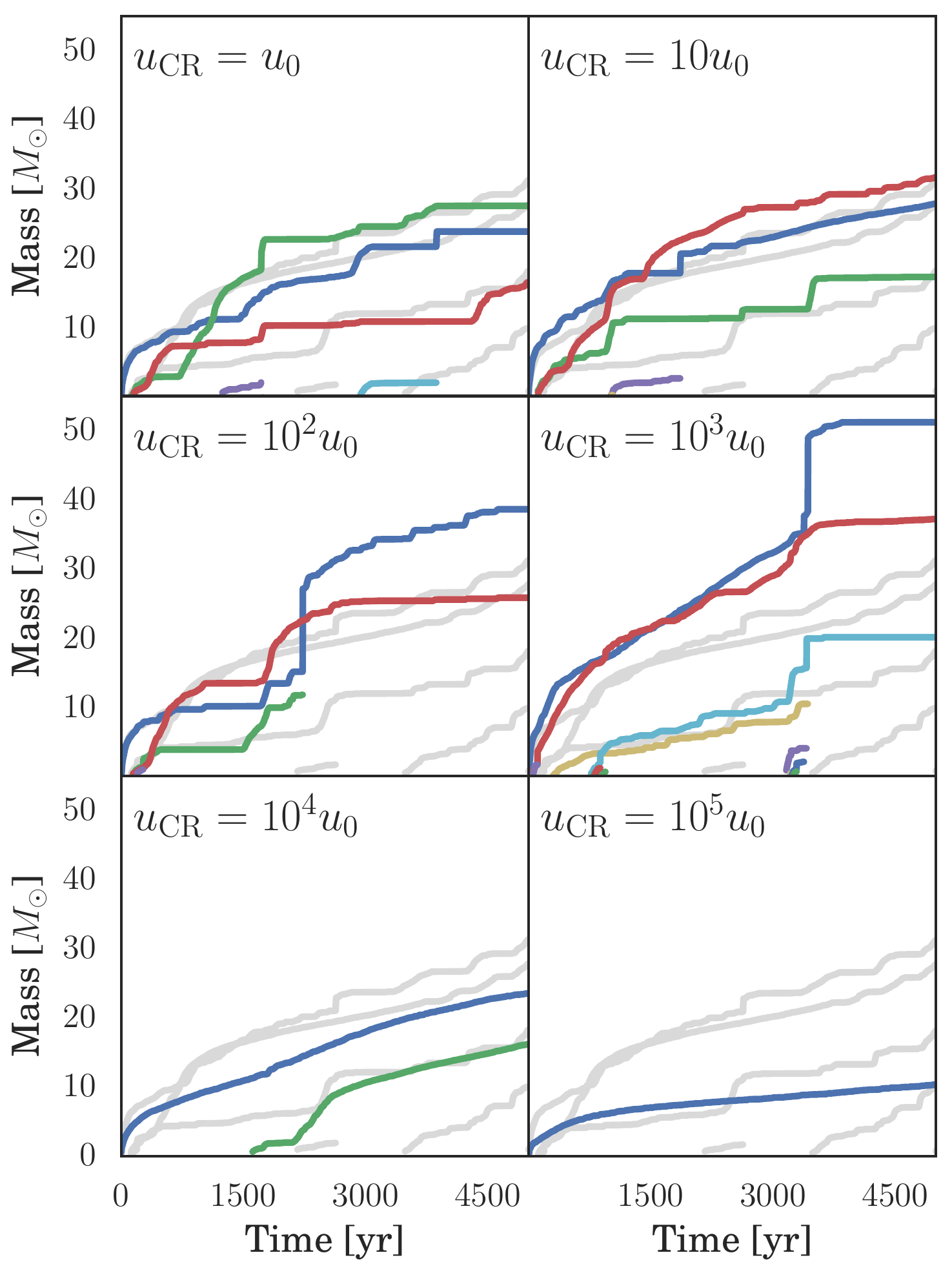}
\caption{\label{fig:sinks} 
Growth of individual sink particles in each simulation over time, with the first sink forming at $t=0$. 
Lines end where one sink is accreted by another. 
The Halo 1 suite of simulations is shown in the left panel, Halo 2 in the right panel.  
The growth of the sinks in the corresponding $\ucr=0$ is shown in grey in the background of each panel for comparison. 
There are no apparent trends with increasing $\ucr$ -- neither in typical sink mass nor in the number of fragments formed.%
}
\end{center}
\end{figure*}

\subsubsection{Expedited Collapse}
\label{sec:expedited_collapse}
Boosting the $\htwo$ fraction in the gas lowers the density threshold for efficient cooling, allowing the gas to fulfil the Rees-Ostriker criterion sooner by driving the cooling time $t_{\rm \small cool}$ below the freefall time $t_{\rm \small ff}$ \citep{ReesOstriker1977}.
This is demonstrated in Figure \ref{fig:efrac}, where we see the density at which the gas begins to cool drop as the CR background strength increases.  
This necessarily expedites the subsequent phases of the collapse, as seen clearly in Figure \ref{fig:collapse}, where we show the total gas mass within 1 pc (physical) of the minihalo's centre as a function of redshift.  
The impact this has on the environment in which the minihalo collapses is seen in Figure \ref{fig:structure}, where we show the final simulation output on various scales for Halo 1 and Halo 2, both in the absence of any CR background ($\ucr=0$) and with $\ucr=10^5\,u_0$. 
Low density filamentary gas is heated and prevented from collapsing, reducing the clumping of the IGM and possibly impacting the early stages of reionisation.
Gas above a $\ucr$-dependent density threshold on the other hand experiences enhanced cooling and thus has its collapse expedited.
As a result, the dynamical environment in which the first sink particles form varies significantly depending on the strength of the CR background.

\begin{figure*}
\begin{center}
\includegraphics[width=1\textwidth]{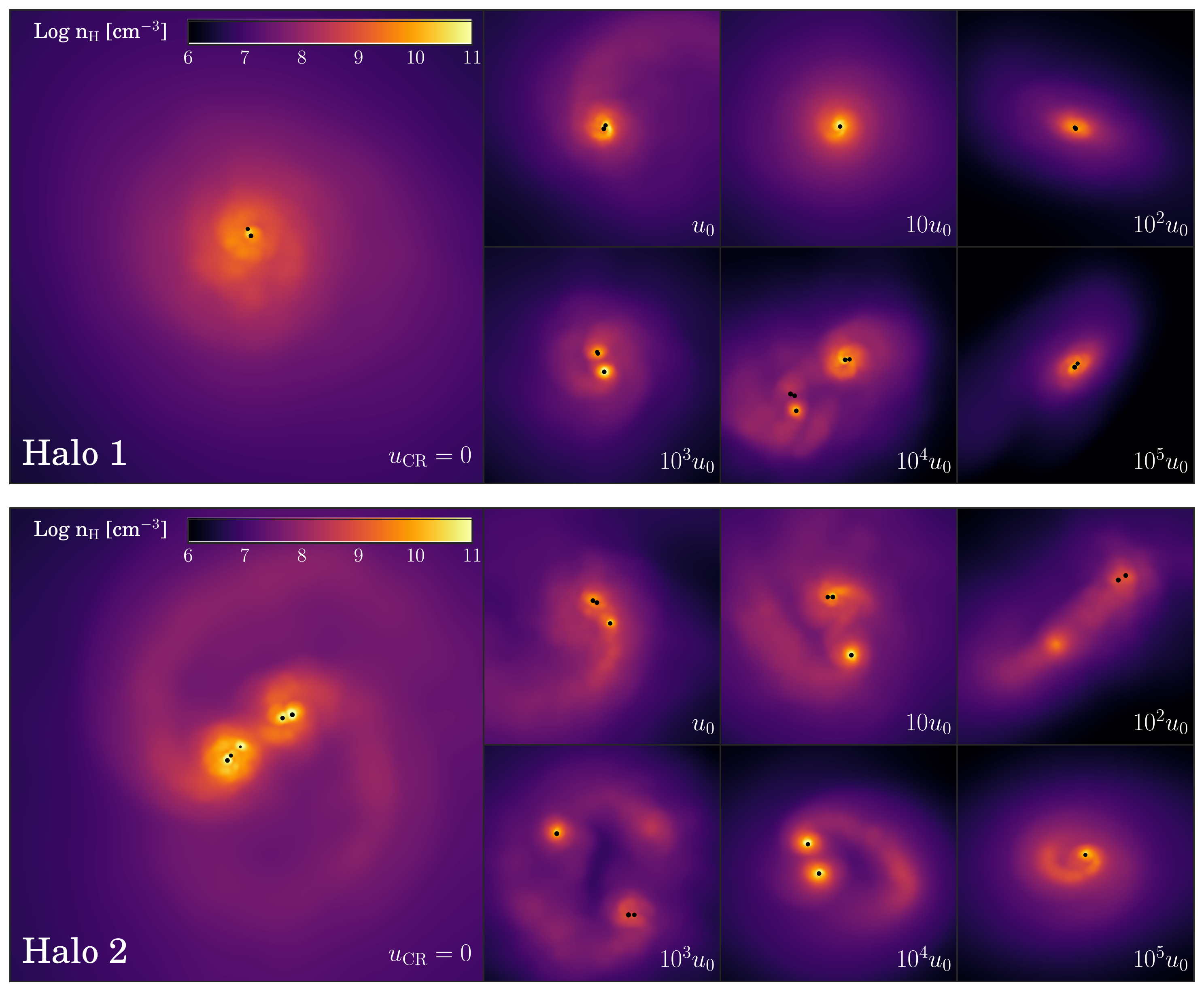}
\caption{\label{fig:disks} 
Density projection of the central $20\,000\au$ of each simulation $5000\yr$ after the formation of the first sink particle.
Upper panel: Halo 1; lower panel: Halo 2.
Each accretion disc is shown face-on; sink particle positions are identified by black dots whose size scales logarithmically with the mass of the sink.
In each panel the accretion disk formed in the absence of any CR background is shown on the left; the corresponding $\ucr$-modified discs are shown to the right.
Note the broad variation in both disc structure and spatial extent.%
}
\end{center}
\end{figure*}

\subsection{Star Formation}
\subsubsection{Protostellar Fragmentation and Growth}
\label{sec:sink_formation}

Figure \ref{fig:sinks} shows the growth over time of all sink particles formed in our simulations, from formation of the first sink particle to simulation's end $5000\yr$ later when radiative feedback can no longer be ignored. 
The first sink particle forms when the gas in the centre of the minihalo reaches densities of $10^{12}\cc$, and develops an accretion disk within a few hundred years. 
In all but three cases, this disk quickly fragments, forming a binary or small multiple within $500\yr$. 
Sinks that survive longer than a few hundred years without undergoing a merger quickly accrete the surrounding gas, growing to $\sim$a few solar masses within $500\yr$ and typically reaching between 10 and $40\msun$ by simulation's end.

As is evident from Figure \ref{fig:sinks}, there is no clear trend with $\ucr$ in either protostellar growth or accretion rate, nor is there any trend in the total sink mass.
For example, both the $10\,u_0$ Halo 1 simulation and $10^5\,u_0$ Halo 2 simulation form only a single sink while the $10^3\,u_0$ Halo 2 simulation steadily forms a total of 14 sink particles over the $5000\yr$ period.
This  can be primarily attributed to the wide variety of dynamical environments in which the sink particles form---demonstrated in Figure \ref{fig:disks}, where we show the accretion disk structure of each simulation $5000\yr$ after the first sink formed.
This variation is a direct consequence of the expedited collapse discussed in Section \ref{sec:expedited_collapse}. 
As seen in Figure \ref{fig:structure}, expediting the initial collapse results in significant variation of the experienced gravitational potential; the collapse history of the gas and the dynamical environment in which the sinks form varies accordingly.
Given the lack of any apparent trends with increasing $\ucr$, this suggests the final stages of the collapse are influenced more by small-scale turbulence than the strength of the CR background.
While CRs may indeed alter the susceptibility of the disk to fragmentation, any such behaviour is masked by variations in the collapse history between simulations.

\begin{figure*}
\begin{center}
\includegraphics[width=1\textwidth]{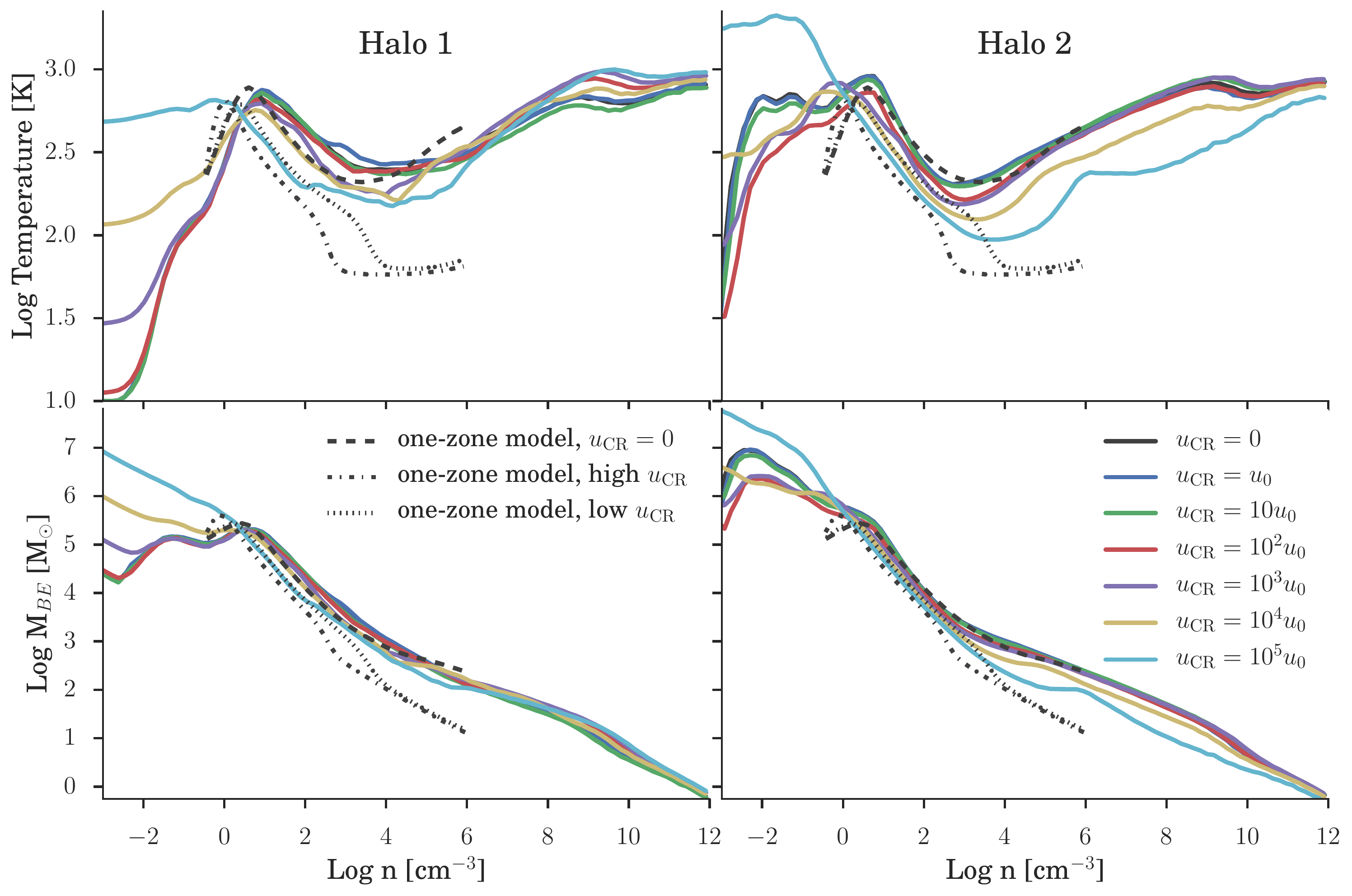}
\caption{\label{fig:Mbe}
Density-binned average temperature (top) and Bonnor-Ebert mass (bottom) for Halo 1 (left) and Halo 2 (right) for all CR background strengths, as labelled.
Overlaid on each panel are the one-zone calculations of \citet{StacyBromm2007}, which only tracked the gas evolution up to densities of $10^6\cc$, insufficient to observe the $\ucr$-independent convergence in the thermodynamic state of the gas as it proceeds to runaway collapse.%
}
\end{center}
\end{figure*}

\subsubsection{Characteristic Mass}

While the final stages of the collapse appear to be somewhat chaotic, with protostellar fragmentation driven primarily by small-scale turbulence rather than the CR background, the typical mass of the sinks formed remains quite stable across all simulations at $10-40\msun$.
This is very much in line with the prevailing consensus for the expected mass of the first stars in the absence of any radiative feedback of $\sim$ a few $\times10\msun$ \citep{Bromm2013}.  
As noted in Section \ref{sec:initial_collapse}, the thermodynamic behaviour of the gas displays a remarkable similarity as it approaches sink formation densities, regardless of $\ucr$.
This suggests that the influence of the CR background is restricted to lower densities, with little to no effect on the protostellar cores from which the first stars ultimately form.

The universality of this characteristic mass is further supported by Figure \ref{fig:Mbe}, where we show the average gas temperature as a function of number density, as well as the approximate fragmentation mass scale, as estimated by the Bonnor-Ebert mass \citep[e.g.,][]{StacyBromm2007}:
\begin{equation}
    M_{\rm \small BE} = 700\msun \left(\frac{T}{200\kelvin}\right)^{3/2}
                                 \left(\frac{n}{10^4\cc}   \right)^{-1/2},
\end{equation}
where $n$ and $T$ are the number density and corresponding average temperature.
Approaching sink formation densities $M_{\rm \small BE}$ is nearly independent of $\ucr$, in agreement with the observed lack of evolution in the sink mass.

Also shown in Figure \ref{fig:Mbe} are the one-zone calculations from \citet{StacyBromm2007}.
Investigating the impact of a CR background on Pop III star formation in a $z=21$ minihalo using the CR background strengths shown in Figure \ref{fig:ucr}, their models suggested 
that a sufficiently strong background might decrease the fragmentation mass scale by an order of magnitude as a result of the cooler temperatures experienced during the loitering phase.
While we observe similar behaviour from gas in the loitering phase, once the gas proceeds to runaway collapse its thermodynamic state quickly converges with that of the $\ucr=0$ case, such that the fragmentation mass scale remains unaltered.
Unfortunately the models of \citet{StacyBromm2007} only followed the gas evolution up to densities of $10^6\cc$, insufficient to observe this convergent behaviour.

\section{Summary and Conclusions}
\label{conclusions}

We have performed two suites of cosmological simulations employing a range of CR backgrounds spanning five orders of magnitude in energy density. 
We follow the thermodynamic evolution of the gas as it collapses into a minihalo from IGM densities up to $n=10^{12}\cc$, at which point three-body processes have turned the gas fully molecular. 
This allows us to capture the combined impact of CR heating and ionisation on Pop III stars forming in the minihalo via its influence on $\htwo$ and $\hd$ cooling.
Once the gas reaches $n=10^{12}\cc$, we form sink particles and follow the subsequent evolution of the system for $5000\yr$, after which point radiative feedback from the forming protostars can no longer be ignored, and our simulations are terminated.

While it also serves to heat the gas, the primary impact of the CR background is to increase the number of free electrons in the collapsing gas, catalysing the formation of additional molecular hydrogen, which in turn enhances the $\htwo$ cooling efficiency.  
This decreases the cooling time, allowing the gas to fulfil the Rees-Ostriker criterion sooner and expediting minihalo collapse. 
Each simulation therefore collapses to high density at a different epoch, and the resulting variation in experienced gravitational potential causes the collapse history---and thus the accretion disk structure---to differ greatly between simulations.

Our first suite of simulations (Halo 1) used the same initial conditions as our prior investigation of a cosmic X-ray background \citep{Hummeletal2015}, allowing for a direct comparison of the results.
CRs are much less efficient than X-rays at heating the gas, such that the collapse suppression observed for a sufficiently strong X-ray background does not occur here.  
While expedited collapse is observed in the presence of an X-ray background as well, the effect is somewhat more pronounced here owing to the more efficient nature of CR ionisation, as there is less associated heating to partially counteract the enhanced cooling.

Our second suite of simulations focused on a minihalo collapsing at $z\simeq21.5$ in the $\ucr=0$ case in order to allow a more direct comparison to the results of \citet{StacyBromm2007}, and control for the influence of the CMB temperature floor on our results.
While we find similar evidence for a lower gas temperature in the loitering phase, extending our study beyond the $n=10^6\cc$ limit of \citet{StacyBromm2007} revealed that the thermodynamic path of the collapsing gas---and thus, the fragmentation mass scale---begins to converge with that of the $\ucr=0$ case for $n\gtrsim10^6\cc$.  
This convergence is observed in all simulations for both Halo 1 and Halo 2, with the thermodynamic state of the gas becoming nearly independent of $\ucr$ by the time we form sink particles at $n=10^{12}\cc$.
The remarkable similarity in the thermodynamic state of the gas just prior to sink formation suggests the presence of a CR background has little impact on the characteristic mass of the stars formed.  
Indeed, the typical mass of the sink particles formed is quite robust, remaining stable across all simulations at $10-40\msun$, very much in line with the expected mass of the first stars in the absence of external radiative feedback. In fact, this robust behaviour is observed under a variety of environmental conditions, including X-ray irradiation \citep{Hummeletal2015} and DM--baryon streaming \citep{StacyBrommLoeb2011a,Greifetal2011b}.

While the neutral impact of the CR background on the thermodynamic state of primordial gas at high densities results in a robust characteristic mass that does not vary with $\ucr$, there is a possibility that the enhanced cooling resulting from CR ionisation may decrease the velocity dispersion of the infalling gas \citep{Clarketal2011a}.  
This would decrease fragmentation in the centre of the minihalo, possibly increasing the mass of the stars formed.
While we see no evidence to support this, any such systematic effect could be easily masked by the variations in collapse history between simulations.  
Likewise, the fact that each simulation collapses to high densities at a different epoch in a different gravitational potential well limits our ability to draw detailed conclusions regarding the impact of the CR background on fragmentation in the protostellar accretion disc.

Finally, while the characteristic mass---and thus, ionising output---of Pop III stars appears to be unaffected by the presence of a CR background, CRs still heat the low-density gas in the IGM, reducing its clumping factor. This may have implications for reionisation and the 21-cm signal \citep{FurlanettoPengBriggs2006, SazonovSunyaev2015}.

\section*{Acknowledgements}
The authors acknowledge the Texas Advanced Computing Center (TACC) at The University of Texas at Austin for providing HPC resources under XSEDE allocation TG-AST120024. This study was supported in part by NSF grant AST-1413501. A.S. gratefully acknowledges support through NSF grant AST-1211729 and by NASA grant NNX13AB84G. This research has made use of NASA's Astrophysics Data System and Astropy, a community-developed core Python package for Astronomy \citep{Robitailleetal2013}.

\bibliography{bibliography/biblio.bib}

\end{document}